\documentstyle[11pt]{article}
\begin{document}

\begin{center}
 {\Large \bf The Spin-Dependent Structure Functions\\[2mm] 
 of Nucleons and Nuclei.\footnote{Talk
 at the Circum-Pan-Pacific Workshop
on High Energy Spin Physics'96, October 2-4, 1996, Kobe, Japan.}}

\vskip .5cm

  \underline{F.C Khanna}$^a$ and {A.Yu.  Umnikov$^b$}

    $^a${\em
    University of Alberta, Edmonton, Alberta T6G 2J1, Canada 
and \\ TRIUMF, 4004 Wesbrook Mall, Vancouver, BC, Canada, V6T 2A3.}

$^b$
{  \em 
INFN, Sezione di Perugia  and 
Department of Physics, University of Perugia,\\
via A. Pascoli, Perugia, I-06100, Italy.} 

\end{center}

 \begin{abstract}
 We discuss 
connection between 
spin-dependent SFs of nucleons
and nuclei in
the  deep inelastic 
lepton scattering. The case of 
the deuteron is studied in detail in the 
Bethe-Salpeter formalism.
 
\end{abstract}

\section{Brief introduction}

In the present talk we discuss the  spin-dependent structure 
functions (SF) of nuclei and their relation to those of nucleon.
Our main focus will be the deuteron, which we study in detail
in the covariant Bethe-Salpeter formalism. 
Why it so important and interesting to study the nuclear effects in the SFs?

First, nuclei are the only source of the experimental information
about  neutron SFs, including the spin-dependent ones.
To obtain this information, it 
is important to understand how nucleons are bound
in the nuclei and how this binding affects 
their SFs. An accurate method to extract
the neutron SFs from the nuclear  data
must be an essential  part of the consistent analysis of the 
nucleon SFs.
Second, a physics governing 
the processes with the participation of the nuclei is extremely interesting 
by itself. 
For instance, a spin-1 nucleus, such as a deuteron, has extra
 spin-dependent
SFs than  nucleons, i.e. $b_{1,2}^D$.
Another example,   nuclei as a slightly relativistic and weakly bound   
systems allows for more progress 
 than the hadrons, in studying the covariant
bound state problem. 
In certain situations the covariant approach gives results
noticeably different from the nonrelativistic ones. 
For the spin-dependent SFs such situation is
a calculation of the $b_{1,2}^D$.
And third, our interest in the study of the reactions with the deuteron
is in part motivated by the future and ongoing experiments. In particular,
very recently we started a study of the chiral-odd SF
$h_1^D$.

\section{Spin-dependent SF of nucleon, $g_1^N$. }

For recent reviews about the nucleon spin-dependent SFs  
 see refs.~\cite{jaffer,rev2}.
\subsection{Basic formulae}

The differential cross section for the polarized electron-nucleon
scattering has the form:
\begin{eqnarray}
\frac{d^2\sigma}{d\Omega dE'} = \frac{\alpha^2E'}{2mq^4E} L^{\mu\nu}W_{\mu\nu},
\label{crosec}
\end{eqnarray}
where  $\alpha = e^2/(4\pi)$, $q=(\nu,0,0,-\sqrt{\nu^2+Q^2})$
 is the momentum transfer,
  $ Q^2=-q^2$, $m$ is the nucleon mass, $E(E')$ is the energy of the incoming (outgoing)
electron, $L^{\mu\nu}$ and $W_{\mu\nu}$ are the leptonic and hadronic tensors.
The most general expression of $W_{\mu\nu}$ is
\begin{eqnarray}
 &&\!\!\!\!\!\!\!\!\!\!\!\!\!\!\!  W_{\mu\nu}^N(q,p) =\label{htenn} \\
&&\!\!\!\!\!\!\!\!\!
 \left ( -g_{\mu\nu} +\frac{q_\mu q_\nu}{q^2}\right ) F_1^N(x_N,Q^2)
 + 
  \left ( p_{\mu} - q_\mu \frac{pq}{q^2}  \right ) 
  \left ( p_{\nu} - q_\nu \frac{pq}{q^2}  \right )
   \frac{F_2^N(x_N,Q^2)}{pq} \nonumber\\
 +&&\!\!\!\!\!\!\!\!\!
\frac{im}{pq} \epsilon_{\mu\nu\alpha\beta} 
   q^\alpha \left \{ S^\beta \left (g_1^N(x_N,Q^2) + g_2^N(x_N,Q^2) \right )
   -p^\beta \frac{(Sq)}{pq}g_2^N(x_N,Q^2) \right \} ,
 \nonumber
 \end{eqnarray}
where
 $x_N = Q^2/(2pq)$ (in the rest frame of the nucleon 
$x_N = Q^2/(2m\nu)$) and 
$S$ is the nucleon spin.

In accordance to the ideology of
the quark-parton model, the SFs $F_{1,2}$ and $g_1$ are
proportional to the appropriate quark distributions on a 
part of the total longitudinal momentum of the nucleon, $x$.
 For instance, if we denote
the net spin carried by quarks as
$\Delta q(x,Q^2)$ ($q = u,d,s$)
and introduce the following combinations:
\begin{eqnarray}
&&\Delta q_3(x,Q^2) \equiv \Delta u(x,Q^2) - \Delta d(x,Q^2) , 
\label{su33}\\
&&\Delta q_8(x,Q^2) \equiv \Delta u(x,Q^2) + \Delta d (x,Q^2) 
-2\Delta s(x,Q^2) ,
\label{su38}\\
&&\Delta \Sigma (x,Q^2) \equiv \Delta u(x,Q^2) + \Delta d(x,Q^2)+ \Delta s(x,Q^2), 
\label{su30}
\end{eqnarray}
then the SFs of the proton(neutron) can be written as:
\begin{eqnarray}
g_1^{p,n}(x,Q^2) = \pm \frac {1}{12}\Delta q_3(x,Q^2) +
\frac{1}{36} \Delta q_8(x,Q^2) +\frac{1}{36}\Delta \Sigma(x,Q^2) .
\label{g1pm}
\end{eqnarray}

Important objects of the study of quark structure of the hadrons 
are the so-called sum rules for the SFs. The sum rules
relate the moments of the SFs to the fundamental 
 (or sometimes not very fundamental) constants  of the theory.
 Integrating eq.~(\ref{g1pm}), the first moments 
of the proton (neutron) structure functions can be written in self-explaining 
notation:
\begin{eqnarray}
S^{p(n)} &\equiv& \int\limits_0^1 dx  g_1^{p(n)}(x,Q^2) =\label{ig1p}\\
&&\!\!\!\! \frac {1}{12}
\left(  1 -\frac{\alpha_s}{\pi}+ {\small \ldots} \right) \left (\pm
\Delta q_3  +
\frac{1}{3} \Delta q_8 \right ) + \frac{1}{9}
\left(  1 -\frac{\alpha_s}{3\pi}+ {\small \ldots} \right)\Delta \Sigma   ,
\label{ignuc}\end{eqnarray}
where the perturbative QCD 
corrections, to order  ${\cal O}(\alpha_s)$, are also presented.

From the current algebra for asymptotic integrals we have ($Q^2 \to \infty $):
\begin{eqnarray}
\Delta q_3  = 1.257\pm 0.003  , \quad
\Delta q_8  = 0.59 \pm 0.02\; (?).
\label{isu38}
\end{eqnarray}
The first constant is from the weak decay of the neutron and 
the second constant is from
the decay of the hyperon.
The ``?'' mark is due to the residual questions about SU(3).

   The Bjorken sum rule is the most fundamental relation:
\begin{eqnarray}
S^p - S^n &=& \frac {1}{6}
\left(  1 -\frac{\alpha_s}{\pi}+ \cdots \right)  \Delta q_3,
\label{bsr}\end{eqnarray}
which numerically gives $0.187 \pm 0.003$ at $Q^2 = 10$~GeV$^2$ and
$0.171 \pm 0.008$ at $Q^2 = 3$~GeV$^2$.

The Ellis-Jaffe sum rule is not so fundamental. Assuming that
$\Delta s = 0$ and, therefore, $\Delta \Sigma =  \Delta q_8 
\simeq 0.6$, we get at $Q^2=10$~GeV$^2$ ($3$~GeV$^2$):
\begin{eqnarray}
S^p_{EJ} &=& \phantom{-}0.171\pm 0.004 \quad 
(\phantom{-}0.161 \pm 0.004) ,\label{ejp}\\
S^n_{EJ} &=& -0.014\pm 0.004 \quad (-0.010 \pm 0.004).\label{ejn}
\end{eqnarray}

The spin-dependent SFs, $g_1$, allow also to study 
spin content of the hadrons. Indeed, using eqs.~(\ref{isu38})
and experimental values of $S^{p,n}$
(a fraction of) the nucleon spin carried by quarks, $\Delta \Sigma$,
can be determined. The total angular
momentum (spin) of the nucleon consists not only of $\Delta\Sigma$, but
also:
\begin{eqnarray}
\frac{1}{2} = \frac{1}{2}\Delta \Sigma + \Delta G + L_z^g +L_z^G, 
\label{spincont}
\end{eqnarray}
where $\Delta G$ is the gluon  spin contribution,
$L_z^{q(G)}$ is the quark (gluon) orbital angular momentum contribution.
In naive quark model $\Delta \Sigma =1$ and others are zeros.
In the relativistic quark model $\Delta \Sigma =0.75$ and $L_z^q=0.125$ 
and
others are zeros.
From the current algebra $\Delta \Sigma \approx 0.6 \pm 0.1$,  others are unknown.

\subsection{Experiments for  $g_1^N$}

Both the SFs and the sum rules are the subject of
intensive experimental studies in recent years.
Table~I presents  measurements by various experimental
groups.

\begin{center}
{\small
{\normalsize \bf Table I.}
\vskip .25cm
\begin{tabular}{|c|c|c|c|c|}
\hline
Experiment & Year & Target      & $\sim Q^2$~GeV$^2$ & $S^{Target} $  \\
\hline
\hline
E80/E130 & 1976/1983 & p         & 5             & 0.17 $\pm$ 0.05 \\
EMC &           1987 & p         & 11            & 0.123 $\pm $ 0.013 $\pm$ 0.019\\
SMC &           1993 & d         & 5             & 0.023 $\pm $ 0.020 $\pm$ 0.015\\
SMC &           1994 & p         & 10            & 0.136 $\pm $ 0.011 $\pm$ 0.011\\
SMC &           1995 & d         & 10            & 0.034 $\pm $ 0.009 $\pm$ 0.006\\
E142 &           1993& n ($^3He$)& 2             & -0.022 $\pm $ 0.011\\
E143 &           1994& p         & 3            & 0.127 $\pm $ 0.004 $\pm$ 0.010\\
E143 &           1995&  d        & 3            & 0.042 $\pm $ 0.003 $\pm$ 0.004\\
HERMES&          1996& n ($^3He$)&   3          & -0.032 $\pm $ 0.013 $\pm$ 0.017\\  
\hline
\end{tabular}
}
\end{center}

\vskip .2cm

From the SMC and E143 data the Bjorken sum rule is:
\begin{eqnarray}
\left (S^p - S^n \right )_{SMC}  &\approx&
0.199 \pm 0.038 \quad{\rm at} \quad Q^2 = 10\quad{\rm GeV}^2,\label{bsre10}\\
\left (S^p - S^n \right) _{E143} &\approx&
0.163 \pm 0.010 \pm 0.016 \quad{\rm at} \quad Q^2 = 3\quad{\rm GeV}^2,\label{bsre3}
\end{eqnarray}
i.e. the sum rule is confirmed with 10 \% accuracy.
From Table~I it is clear that the Ellis-Jaffe sum rules are broken.

As to the spin content, (\ref{spincont}), only one piece, $\Delta \Sigma$, 
can be extracted from the 
data for the integrals of SFs. 
The world data from   Table~I
gives:
\begin{eqnarray}
\Delta \Sigma \approx 0.3\pm 0.1,
\label{spinconte}
\end{eqnarray}
which is larger than the first result of EMC, 
$\Delta \Sigma = 0.12\pm 0.094\pm 0.138\approx 0$, but still 
lower than quark model estimates.

In addition to the perturbative corrections in eq.~(\ref{ig1p}),
various other corrections, such as
the kinematic
mass corrections, $\sim m^2/Q^2$ and higher twist  corrections, $\sim 1/Q^2$,
 are discussed.

\section{Nucleons and nuclei}

Note that actual data for the neutron is not presented
in  Table~I, only the data for lightest nuclei.
A simple formula is used to obtain $g_1^n$ from the combined 
proton and deuteron
data:
\begin{eqnarray}
g_1^D = \left (1-\frac{3}{2}w_D\right )\left (g_1^p+g_1^n\right),
\label{simple}
\end{eqnarray}
where $w_D$ is the probability of the $D$-wave state in the deuteron.
Depending on the model,
$w_D = 0.04 - 0.06$.
Similarly, the neutron SF is obtained 
from the $^3He$ data:
\begin{eqnarray}
g_1^{^3He} = \left ( P_S + \frac{1}{3}P_{S'} -P_D
\right  )
g_1^n
+\left (\frac{2}{3}P_{S'} - \frac{2}{3}P_D
\right  )
g_1^p,
\label{simple1}
\end{eqnarray}
where $P_S$, $P_{S'}$ and $P_D$ are weights of the $S$, $S'$ and 
$D$ waves in $^3He$, respectively. Typical (model-dependent) 
values of these weights
are $P_S\approx 0.897$, $P_{S'}=0.017$ and $P_D=0.086$.

It is important to realize that the real connection between the nucleon and 
nuclear SFs is more complex than given by formulae like
eq.~(\ref{simple}) and (\ref{simple1}). Studies of the last decade 
show importance of the proper separation
of the binding, Fermi motion  and the off-mass-shell 
effects in the procedure of extracting the neutron SFs from the 
nuclear data
(see refs.~\cite{amb,unfo,mtamb} and references therein).
However, effects of the Fermi motion are sometimes 
estimated by the experimental groups, other effects are always 
neglected.
Such a way of action can be phenomenologically more or less safe
at the present level of accuracy of the experiments, but 
not in general.

 The deuteron is the most appropriate target to study the neutron
SFs, since it has a well-known structure and
well-studied wave function or relativistic amplitude. Besides 
all other effects such as meson exchanges, binding of the nucleons,
off-mass-shell corrections, shadowing, etc, are minimal.
Even in the case of  $^3$He the situation is
known to be different~\cite{kubj,fshe}. Indeed, 
eq.~(\ref{simple1}) or even more sophisticated convolution formula
violate the fundamental Bjorken sum rule for the 
$^3$He-$^3$H pair at the 3-5\% level, which is a serious
indication of other degrees of freedom involved in the process.
Once again, this fact is completely ignored by the experimental
groups reporting the results for the neutron SFs
from experiments with $^3$He.
In what follows we consider the nuclear effects in the
spin-dependent
SFs  of the deuteron.
Results of our studies make us certain that an accurate 
extraction of the {\em neutron } spin structure
function, $g_1^n$,
is possible.

Considering nuclei as a complex system of interacting nucleons and mesons,
we calculate the nuclear SFs in terms of the structure
functions of its constituents, nucleons and mesons, and in 
the Bethe-Salpeter formalism for the deuteron amplitude. 
For the spin-independent SFs, $F^D_2$, the mesonic
contributions to the SF is important
(although quite small) for the consistency
of the approach, since the mesons carry a part of the total momentum 
of the nuclei (see~\cite{uk} and references therein).
 However, for the spin-dependent
SFs explicit contribution of mesons 
is not important. Rather,
their presence manifests  via binding of nucleons in nuclei. 
This is why we consider only nucleon contributions   to the 
spin-dependent
SFs.

We start with the general form of the hadron tensor 
of the deuteron with the total angular momentum projection, $M$,
keeping  only 
leading twist SFs:
\begin{eqnarray}
  W_{\mu\nu}^D(q,P_D,M) &=&  \left ( -g_{\mu\nu} +\frac{q_\mu q_\nu}{q^2}\right ) F_1^D(x_D,Q^2,M)
 + \label{htend} \\
&& \left ( P_{D\mu} - q_\mu \frac{P_Dq}{q^2}  \right ) 
  \left ( P_{D\nu} - q_\nu \frac{P_Dq}{q^2}  \right )
   \frac{F_2^D(x_D,Q^2,M)}{P_Dq} \nonumber\\
  && +\frac{iM_D}{P_Dq} \epsilon_{\mu\nu\alpha\beta} 
   q^\alpha S_D^\beta(M) g_1^D(x_D,Q^2),
 \nonumber
 \end{eqnarray}
where
  $x_D = Q^2/(2P_Dq)$ (in the rest frame of the deuteron 
$x_D = Q^2/(2M_D\nu)$), 
$S_D(M)$ is the deuteron spin 
 and $F_{1,2}^D$ and $g_1^D$
are the deuteron SFs. 
Averaged over  $M$ this expression leads to the well-known form of the
spin-independent hadron tensor which is valid for hadron with any spin:
{\small\begin{eqnarray}
 &&  W_{\mu\nu}^D(q,P_D) = \frac{1}{3}\sum_M W_{\mu\nu}^D(q,P_D,M) \label{av}\\
   &=&\!\!\!\!
 \left ( -g_{\mu\nu} +\frac{q_\mu q_\nu}{q^2}\right ) F_1^D(x_D,Q^2)
 + 
  \left ( P_{D\mu} - q_\mu \frac{P_Dq}{q^2}  \right ) 
  \left ( P_{D\nu} - q_\nu \frac{P_Dq}{q^2}  \right )
   \frac{F_2^D(x_D,Q^2)}{P_Dq},
 \nonumber
 \end{eqnarray}
}
where $F_{1,2}^D(x_D,Q^2)$ are the result of averaging of the SF 
$F_{1,2}^D(x_D,Q^2,M)$.

To  separate $g_1^D$ we can use one of the  following projectors~\cite{ukk}:
 \begin{eqnarray}
  R^{(1)}_{\mu\nu} \equiv i\epsilon_{\mu\nu\alpha\beta}q^\alpha S^\beta_D(M),
\quad R^{(2)}_{\mu\nu} \equiv \frac{i (S_D(M)q)}{P_Dq}
\epsilon_{\mu\nu\alpha\beta}q^\alpha P_D^\beta
. \label{aw2}
\end{eqnarray}
In the limit $Q^2/\nu^2 \to 0$:
 \begin{eqnarray}
g_1^D = \frac{R^{(1)\mu\nu}W_{\mu\nu}^D }{2\nu} =
\frac{R^{(2)\mu\nu}W_{\mu\nu}^D }{2\nu}
. \label{exg1}
\end{eqnarray}

The nucleon contribution to the deuteron SFs 
is presented by the triangle graph, written in terms of the Bethe-Salpeter
amplitude of the deuteron~\cite{uk,ukk}:

\let\picnaturalsize=N
\def\picsize{2.10in}
\def\picfilename{tre-dia.eps}
\ifx\nopictures Y\else{\ifx\epsfloaded Y\else\input epsf \fi
\let\epsfloaded=Y
\centerline{\ifx\picnaturalsize N\epsfxsize \picsize\fi \epsfbox{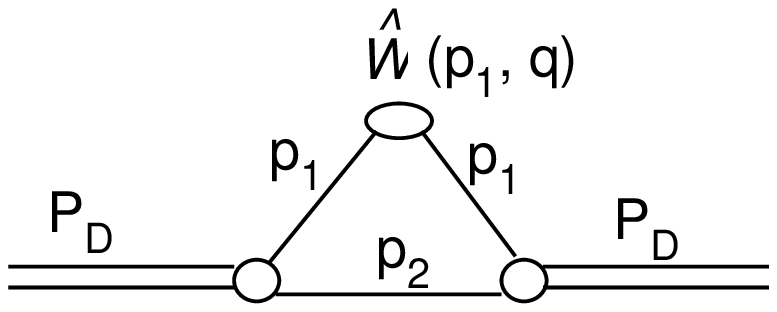}}}\fi
\noindent
where $\hat W$ is the appropriate operator, describing the scattering on the
constituent nucleon. Neglecting small correction due to the ``nucleon 
deformation''~\cite{mst} it can be written down as:
\begin{eqnarray}
&&\hat W^N_{\mu\nu}(q,p)=\hat W_{\{\mu\nu\}}(q,p) + 
\hat W_{[\mu\nu]}(q,p) \label{deftens}\\[3mm]
&&\hat W_{\{\mu\nu\}}(q,p) = \frac{\hat q}{2pq}W^N_{\mu\nu}(q,p),
 \label{ops}\\
&&\hat W_{[\mu\nu]}(q,p) = 
\frac{ i}{2pq}
\epsilon_{\mu\nu\alpha\beta}
q^{\alpha}
\gamma^\beta \gamma_5 g_1^N (q,p)
,
 \label{opa}
\end{eqnarray}
where $\{\ldots \}$ and $[\ldots ]$ denote symmetrization and 
antisymmetrization of indices, respectively, $W_{\{\mu\nu\}}^N$
is the spin-independent part of the hadron tensor of the nucleon
 and $g_1^N(q,p)=g_1^N(x,Q^2)$ is the spin-dependent
nucleon  SF.

The explicit expressions of the deuteron SFs in 
terms of the Bethe-Salpeter amplitude, $\Psi_M(p_0,{\bf p})$, are given by:
\begin{eqnarray}
 F_2^D(x_N,Q^2,M) &=& i\int \frac{d^4p}{(2\pi)^4}
{F_2^N} \left( \frac{x_N m}{p_{10}+p_{13}}, Q^2\right)\label{f2m} \\
&&\frac{ {\sf Tr}\left\{
\bar\Psi_M(p_0,{\bf p})(\gamma_0+\gamma_3) \Psi_M(p_0,{\bf p}) (\hat p_2-m)
\right \}}{2M_D},
\nonumber \\[2mm]
 g_1^D(x_N,Q^2) &=& i
  \int \frac{d^4p}{(2\pi)^4}
{g_1^N}\left( \frac{x_N m}{p_{10}+p_{13}}, Q^2\right) \label{g1m}\\
&& \frac{\left. {\sf Tr}\left\{
\bar\Psi_M(p_0,{\bf p})(\gamma_0+\gamma_3)\gamma_5 \Psi_M(p_0,{\bf p}) (\hat p_2-m)
\right \}\right |_{M=1}}{2(p_{10}+p_{13})},
 \nonumber 
\end{eqnarray}
where 
 $p_{10}$ and $p_{13}$ are the time and 3-rd components of the
struck nucleon momentum. 
Averaging over the projection  $M$ has not been done
in eq.~(\ref{f2m}),
since we use  the present form later to calculate
the  SF $b_{1,2}^D$. Then  two
independent ``SFs'', with $M = \pm 1$ and $M=0$ are obtained:
\begin{eqnarray}
&&F_2^D(x_N,Q^2) = \frac{1}{3} \sum_{M=0,\pm 1} F_2^D(x_N,Q^2,M),
 \label{f2}\\[1mm]
&& F_2^D(x_N,Q^2,M=+1) = F_2^D(x_N,Q^2,M=-1).\label{pm}
\end{eqnarray}

A method to calculate numerically   expressions like 
(\ref{f2m}) and (\ref{g1m}) 
is presented in ref.~\cite{ukk}. The important details of
the calculations are:
\begin{enumerate}
\item A realistic model for the Bethe-Salpeter amplitudes 
is essential for a realistic estimate of the 
nuclear effects. We use a recent numerical solution~\cite{uk} 
of the ladder Bethe-Salpeter equation with a realistic  
exchange kernel. 
\item The Bethe-Salpeter amplitudes and, therefore, 
eqs.~(\ref{f2m})-(\ref{g1m})
have a nontrivial singular structure. These singularities 
must be  carefully taken into account.
\item The BS amplitudes are numerically calculated with the
help of the Wick rotation. Therefore, the  numerical
procedure for  inverse Wick rotation must be applied.
\end{enumerate}

\begin{center}
 \vspace*{-5cm}
\let\picnaturalsize=N
\def\picsize{4.8in}
\def\picfilename{g1.ps}
\ifx\nopictures Y\else{\ifx\epsfloaded Y\else\input epsf \fi
\let\epsfloaded=Y
\centerline{\ifx\picnaturalsize N\epsfxsize \picsize\fi \epsfbox{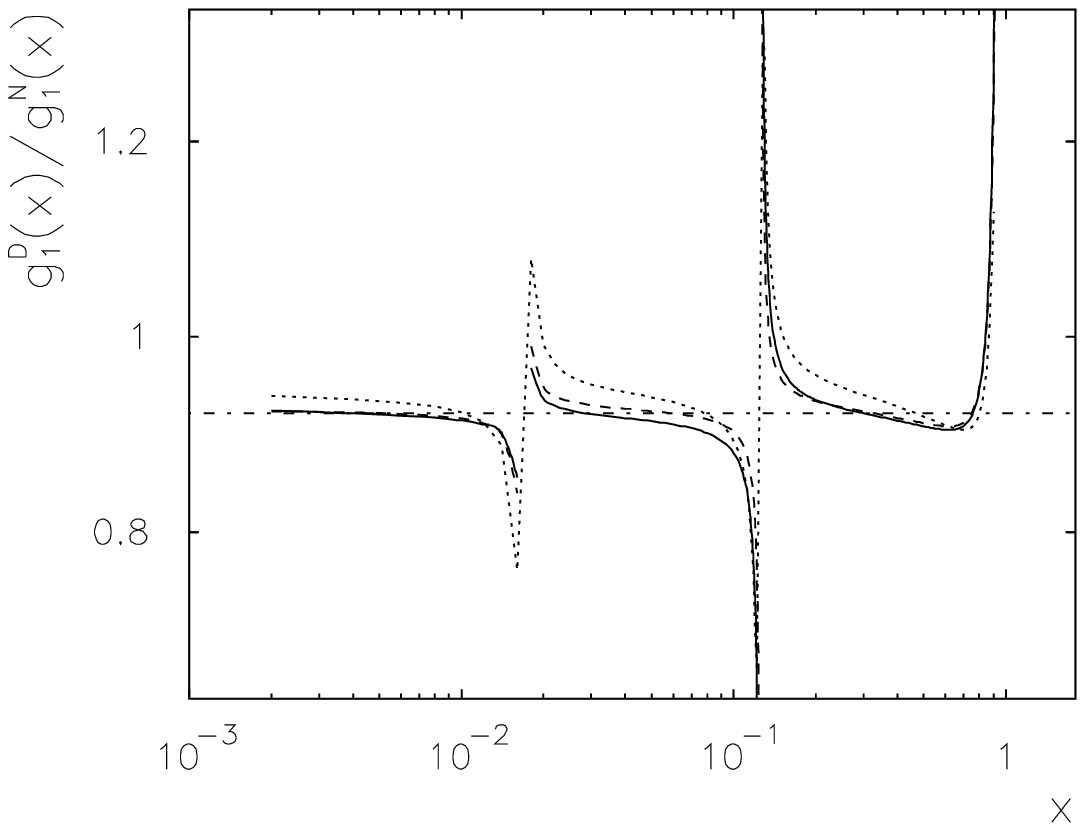}}}\fi

\vspace*{-5.5cm}
Figure~1. 
\end{center}

Calculations with the realistic BS amplitudes result in the 
behavior of the $g_1^D$ very similar to 
other calculations~\cite{kaemp,mg1}. The result is presented in Fig.~1
in the form of the ratio, $g^D_1/g_1^N$. Dotted curve presents 
non-relativistic calculations with the Bonn wave function, solid curve the 
BS result. The dashed curve is the illustrative result for the 
non-relativistic calculations   utilizing the BS axial-vector 
density. Last curve, dot-dased, in   Fig.~1 corresponds to the 
naive formula (\ref{simple}).
Despite seemingly drastic difference in the ratio $g^D_1/g_1^N$ given by
(\ref{simple}) and realistic calculations, the typical experimental 
errors today are larger. (Large fluctuations of the ratio 
 at $x<0.7$ are not too important. They correspond to  zeros
 of the nucleon SF which are slightly shifted by the convolution formula.)
 Indeed, in Fig.~2 we present representative example of data
 (SMC-1994), together with two fits of these data (dashed lines).
 The solid lines  present results of {\em exact} 
 extraction of the nucleon SF from the present deuteron data.
 We see that curves for deuteron and nucleon both do not contradict 
 the experiment.

However, in certain kinematical conditions effects can be  bigger.
For instance, lately much interest is   devoted to the discussion
of the Gerasimov-Drell-Hearn sum rule for the proton and the neutron
SFs  at small $Q^2$ in general and
at $Q^2 = 0$ in particular (see reviews~\cite{jaffer,rev2}
and references therein). Very important   contribution
to the study of the neutron SFs
is expected from  the Jefferson Lab groups~\cite{expgdh}, where
experiments with the deuterons and $^3He$ are planned in the intervals
$Q^2\sim 0.15 - 2$~GeV$^2$.
Analysis of the deuteron SFs in this interval of $Q^2$,
the nucleon ``resonances'' region, shows that effect of the binding and Fermi
motion is much larger here than in the deep inelastic regime.
An example of the calculation of the deuteron structure 
function, $g_1^D(x,Q^2)$, is presented in Fig.~3a (dashed line)
at $Q^2=1.0$~GeV$^2$. It is compared 
with the nucleon SF, $g_1^N(x,Q^2)$, input into the calculation.
In the areas of resonance structures in  $g_1^N(x,Q^2)$, the deuteron
SF differs  up to 50\%!
In Fig.~3b we present a comparison of the neutron SF,
$g_1^n$ (solid line, input into calculations in Fig.~3a),
 with the ``neutron''
SF ``extracted'' by means of the naive formula (\ref{simple})
(dashed line).
We see that these  two  functions have nothing in common.

\vspace*{-2mm}
\begin{center}
\let\picnaturalsize=N
\def\picsize{4.0in}
\def\picfilename{g1-extract.eps}
\ifx\nopictures Y\else{\ifx\epsfloaded Y\else\input epsf \fi
\let\epsfloaded=Y
\centerline{\ifx\picnaturalsize N\epsfxsize \picsize\fi \epsfbox{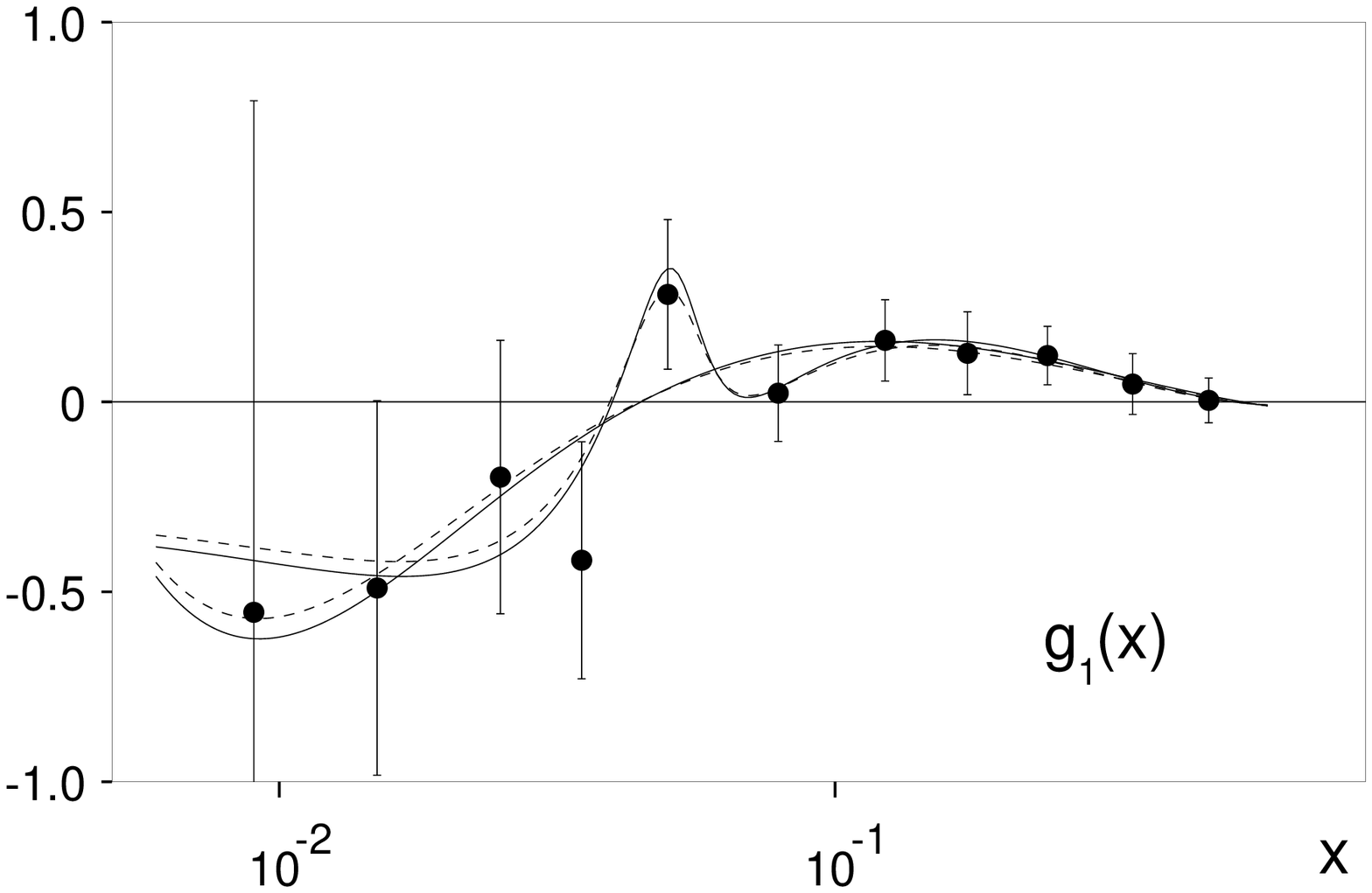}}}\fi
Figure~2. 
\end{center}

The same effects appear in  $^3He$~\cite{ciofi}.
The presented example, shows that in every particular situation
one has to consider the nuclear effects and take into account 
corresponding corrections to the SFs.
 
Mathematically the problem of extraction of the neutron
SF from the deuteron data is formulated as a problem to solve
the inhomogeneous integral equation (\ref{g1m})
for the neutron SF with a model kernel 
and experimentally measured left hand side\footnote{Depending
on the model, some
additive corrections could be taken into account.}, $g_1^D$. 
Recently we proposed a method 
to extract the
neutron SF from the deuteron data
within any 
model, giving deuteron SF in the form
of a "convolution integral plus/minus additive corrections"~\cite{unfo}. 
The principal advantages of the method, compared with 
the smearing factor method, are the following. 
(i) Only analyticity of the SF need 
be assumed, (ii) the
method allows us to elaborate on the spin-dependent SF,
where the traditional smearing factor method does not work.

\begin{center}

\begin{minipage}{15cm}

\vspace*{-7.05cm}

\hspace*{-2.5cm}
\let\picnaturalsize=N
\def\picsize{3.60in}
\def\picfilename{gdh-sf1.ps}
\ifx\nopictures Y\else{\ifx\epsfloaded Y\else\input epsf \fi
\let\epsfloaded=Y
\centerline{\ifx\picnaturalsize N\epsfxsize \picsize\fi \epsfbox{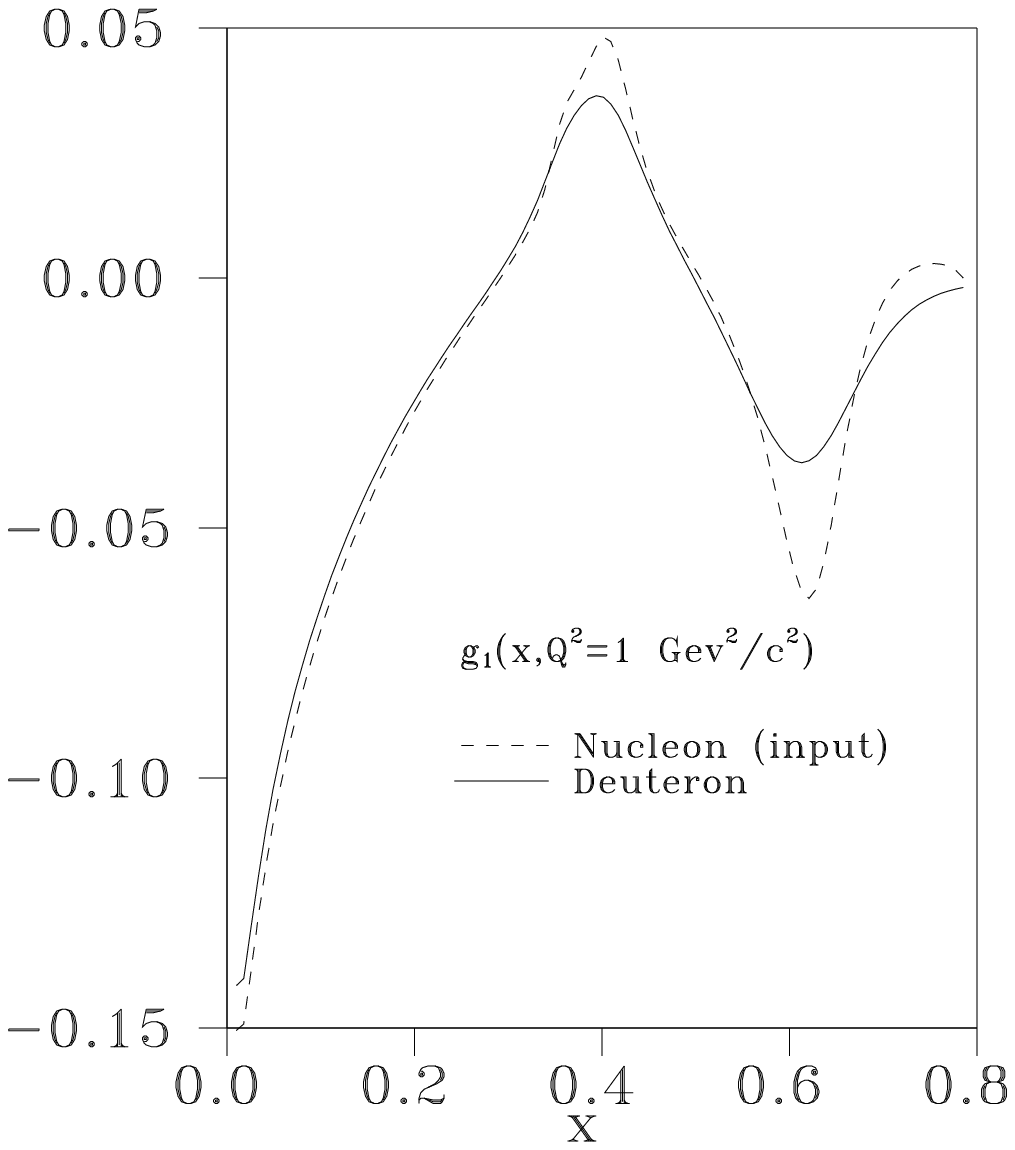}}}\fi
 
\vspace*{-12.7cm} 

\hspace*{4.5cm}
\let\picnaturalsize=N
\def\picsize{3.60in}
\def\picfilename{gdh-ntr1.ps}
\ifx\nopictures Y\else{\ifx\epsfloaded Y\else\input epsf \fi
\let\epsfloaded=Y
\centerline{\ifx\picnaturalsize N\epsfxsize \picsize\fi \epsfbox{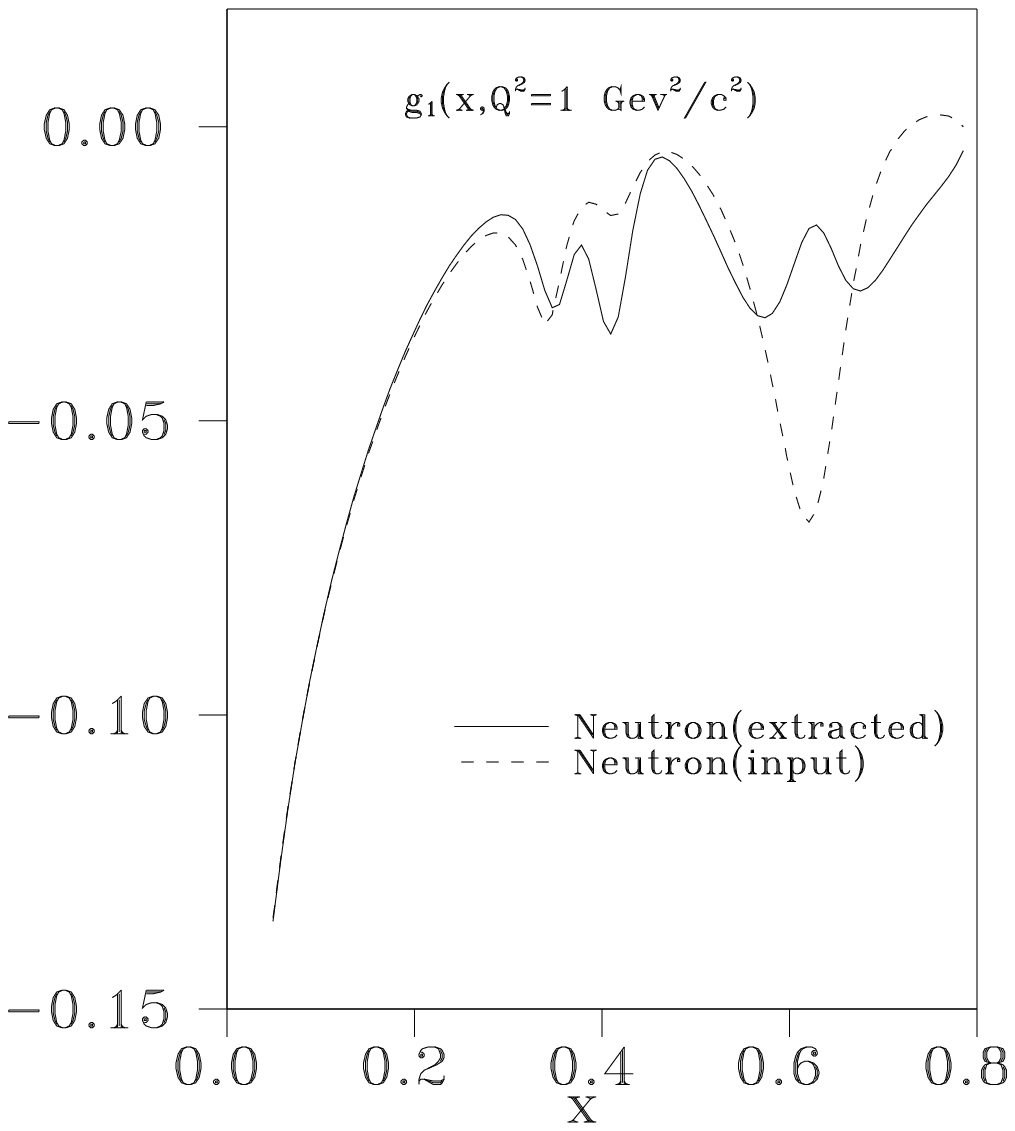}}}\fi
\end{minipage}
\vskip .9cm

a)  \hspace*{6cm} b)

Figure~3.
\end{center}

\section{Other spin-dependent structure functions}

\subsection{SFs for spin-1 hadron, ${b_{1,2}^D}$}.

The SF $b_1^D$ is defined by
(see ref.~\cite{ukk,ub} and references therein):
\begin{eqnarray}
&& b_2(x_N,Q^2) = F_2^D(x,Q^2,M=+1)-F_2^D(x,Q^2,M=0),\label{b2}   
\end{eqnarray}

Note, the SF $F_2^D(x,Q^2,M)$ is independent of the 
lepton polarization, therefore, both  SFs, $F_2^D$ 
and $b_2^D$, can be measured in  experiments with an unpolarized
lepton beam and polarized deuteron target. In view of  eq.~(\ref{pm}),
only one  of the  SFs $F_2^D(x,Q^2,M)$ is needed, in addition
to the spin-independent $F_2^D(x,Q^2)$, in order to obtain
$b_2(x,Q^2)$. The other  SF, $b_1^D$, is related
to the deuteron  SF $F_1^D$, the same way as $b_2^D$ is
related to $F_2^D$,  via eqs.~(\ref{f2}), and $b_2^D = 2xb_1^D$.

Sum rules for the deuteron  SFs $b_1^D$ and $b_2^D$
are a 
result of the fact that the vector charge and energy of the system are
independent of the spin orientation:
\begin{eqnarray}
\int\limits_0^1 dx_D b_1^D(x_D) =0, \quad
\int\limits_0^1 dx_D b_2^D(x_D) =0.
 \label{sr2}
\end{eqnarray}
These sum rules 
were suggested by Efremov and Teryaev~\cite{et}.

\vspace*{-2.75cm}

\begin{center}
\let\picnaturalsize=N
\def\picsize{3.90in}
\def\picfilename{b2.ps}
\ifx\nopictures Y\else{\ifx\epsfloaded Y\else\input epsf \fi
\let\epsfloaded=Y
\centerline{\ifx\picnaturalsize N\epsfxsize \picsize\fi \epsfbox{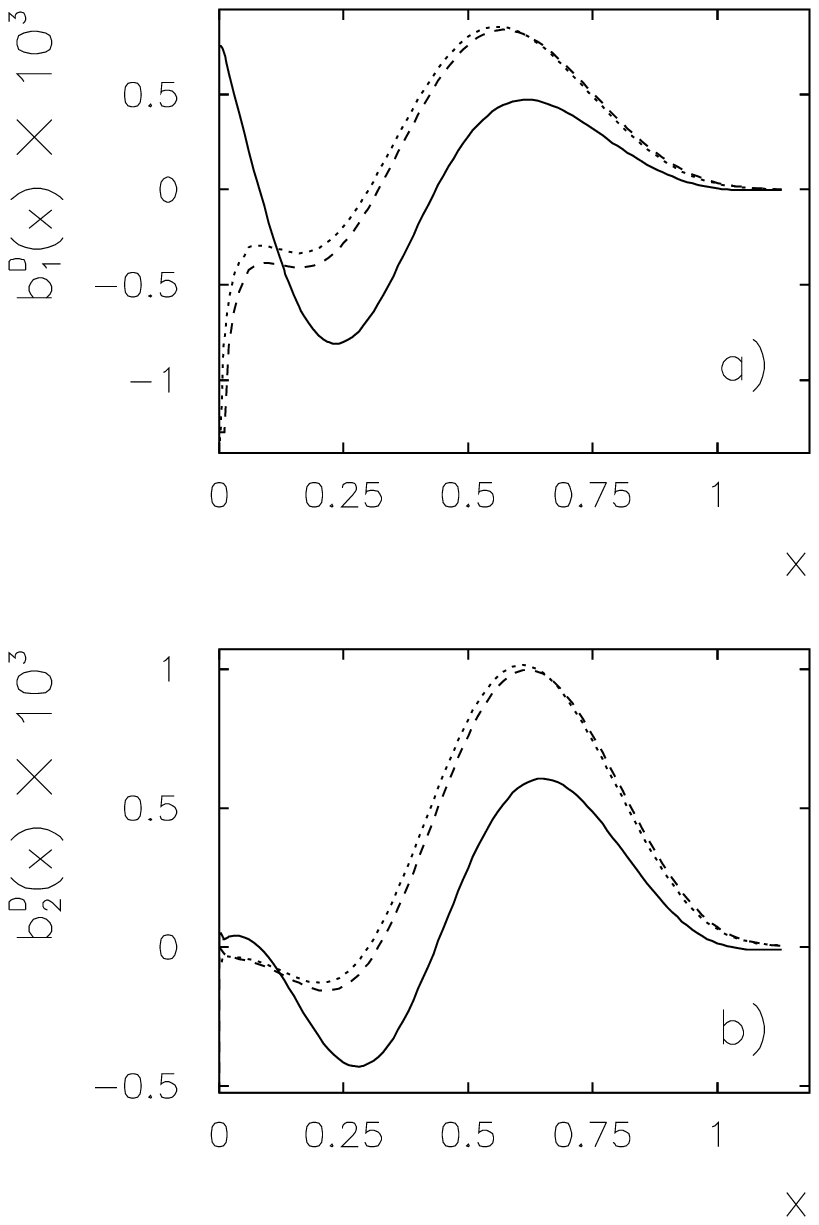}}}\fi

\vspace*{-3.2cm}
Figure~4.
\end{center}

The  SFs $b_1^D$ and $b_2^D$ are calculated within 
two approaches as well. The results are shown in Fig.~4 a) and b).
The behavior of the functions in Fig.~4 a) suggests the validity of the 
first of sum rules (\ref{sr2}). At the same time,  the nonrelativistic
calculation for 
$b_2^D$ in  Fig.~4 b) (dotted line) obviously
does not satisfy the second sum rule. The main difference of
 the relativistic and nonrelativistic calculations is at small $x$, where
 these approaches give different signs for the  SFs. To check a model
 dependence of  the nonrelativistic calculations, we also performed
 calculations with the
  ``softer'' deuteron wave function (with cut-off of the realistic 
 wave function at $|{\bf p}| = 0.7$~GeV). 
 Corresponding  SFs are shown in Fig.~4 a) and b) (dashed line).
 It also does not affect the principle conclusion that the nonrelativistic
 approach violates the sum rules.

\subsection{Chiral-odd SF ${h_{1}^D}$}.

The spin-dependent SFs $h_1$ of the nucleons and deuteron
can not be measured in the inclusive deep inelastic
scattering, but in the semi-inclusive process~\cite{jj}. In this sense
these SFs are  different from
the SFs studied in the present paper.
However, we present the results for these functions,
since (i) they carry important information about the spin structure
of the nucleons~\cite{jj,hj} and the deuteron~\cite{uhk},
(ii) the experiments are planned to measure them~\cite{exp}
and (iii) from the theoretical point of view structure 
function of the deuteron, $h_1^D$, is defined in a way very
similar to the usual deep inelastic SFs~\cite{uhk}:
\begin{eqnarray}
 h_1^D(x_N) &=& i
  \int \frac{d^4p}{(2\pi)^4}
{h_1^N}\left( \frac{x_N m}{p_{10}+p_{13}}\right) \label{h1m}\\
&&\frac{\left. {\sf Tr}\left\{
\bar\Psi_M(p_0,{\bf p})\gamma_5\gamma_3\gamma_0 \Psi_M(p_0,{\bf p})
 (\hat p_2-m)
\right \}\right |_{M=1}}{2(p_{10}+p_{13})}.
 \nonumber
\end{eqnarray}

To calculate the realistic SF $h_1^D(x)$ we need
 the nucleon SFs $h_1^N(x)$. However, so far there is 
no existing experimental data for this function, and very little is known
about the form of $h_1^N$ in theory.  
In the present paper
we follow the ideas  of ref.~\cite{jj} to estimate $h_1^N$. 
Since the sea quarks do not contribute to $h_1^N$, its 
flavor content is
simple:
\begin{eqnarray}
h_1^N (x)  = \delta u(x) +\delta d(x), 
\label{hn1}
\end{eqnarray}
where $\delta u(x) $ and $\delta d(x)$ are the contributions 
 of the u-
and d-quarks, respectively~\cite{jj,hj}.
Since the matrix elements of the operators $\propto \gamma_5 \gamma_3$
and
$\propto \gamma_5 \gamma_3\gamma_0$ coincide in the static limit,
 as a crude estimate we can expect that
\begin{eqnarray}
\delta u(x) \sim \Delta u(x), \quad \quad 
\delta d(x) \sim \Delta d(x), \label{dud1}
\end{eqnarray}
where $\Delta u(x)$ and $\Delta d(x)$ are contributions 
of the u- and d-quarks to the spin of the nucleon, which is
measured through the
SF $g_1^N$. Correspondly, the simplest estimation
for $h_1^N$
\begin{eqnarray}
h_1^N (x)  = \alpha \Delta u(x) +\beta \Delta d(x), 
\label{hn2}\\
\alpha = \beta=1 \label{ab1}
\end{eqnarray}
should not be too unrealistic. In fact, 
the bag model calculation shows that difference between $\delta q$ and 
$\Delta q$ is typically only few percent~\cite{jj}. This analysis is
mostly a qualitative one, since it is 
limited by the case with one quark flavor and does not pretend to describe
a phenomenology.

To evaluate possible deviations from the simple  choice 
of $h_1^N$, (\ref{hn2}) with (\ref{ab1}), we
suggest:
\begin{eqnarray}
\alpha = \delta u/\Delta u, \quad \beta= \delta d/\Delta d, \label{ab2}
\end{eqnarray}
where  $\delta q$  and $ \Delta q$  are the first moments of 
$\delta q(x)$  and $ \Delta q(x)$, respectively ($q = u, d$).
For 
$\delta u$  and $ \delta d$ we can adopt the results from the QCD sum rules 
and the bag model calculations~\cite{hj}. As to
$\Delta u$  and $ \Delta d$, we 
can use  the experimental data analysis~\cite{jaffer,rev2}
or theoretical results, e.g.
the QCD sum rule results~\cite{hj}.
Thus, we estimate~\cite{uhk}:
\begin{eqnarray}
\alpha = 1.5 \pm 0.5, \quad \beta= 0.5 \pm 0.5, \label{ab3}
\end{eqnarray}
at the scale of $Q^2=1$~GeV$^2$.

\vskip -.7cm

\begin{center}
\let\picnaturalsize=N
\def\picsize{3.0in}
\def\picfilename{h-1d.eps}
\ifx\nopictures Y\else{\ifx\epsfloaded Y\else\input epsf \fi
\let\epsfloaded=Y
\centerline{\ifx\picnaturalsize N\epsfxsize \picsize\fi \epsfbox{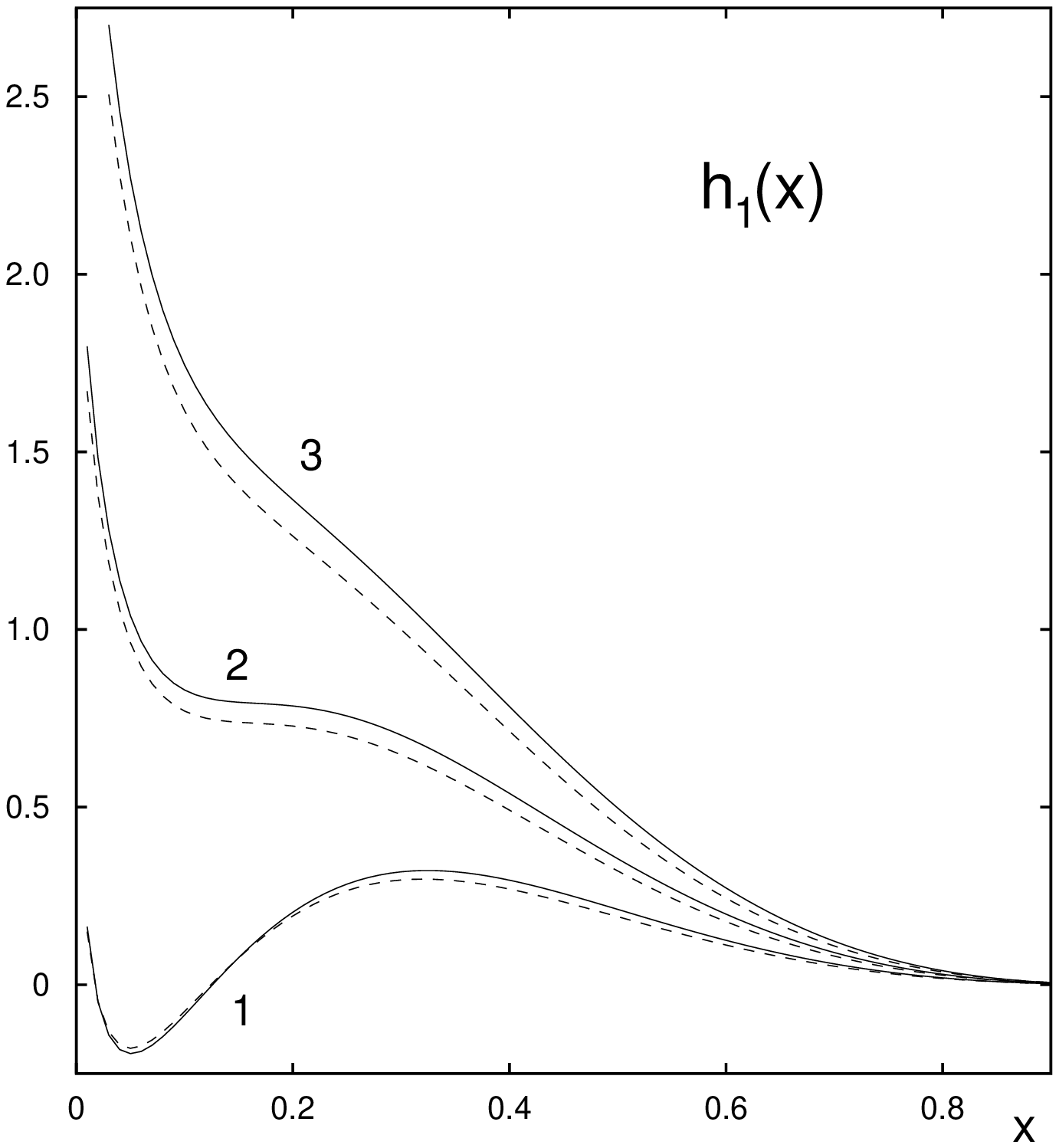}}}\fi
Figure~5.
\end{center}

The realistic form of the distributions $\Delta u(x)$ and $\Delta d(x)$ 
can be taken from a fit to the experimental data for $g_1^N$. In our
calculations we used parametrization from ref.~\cite{shaf}. 
At this point we have to realize that, in spite of expected 
relations~(\ref{dud1}), distributions $\delta q$ and 
$\Delta q$ are very different in their nature. Especially at  
$x {\ \lower-1.2pt\vbox{\hbox{\rlap{$<$}\lower5pt\vbox{\hbox{$\sim$}}}}\ }0.1$, where $\Delta q$ probably contains a singular contribution
of the polarized sea quarks, but $\delta q$ does not.
Therefore we expect eq.~(\ref{hn2}) to be a reasonable 
estimate in the region of the valence quarks  dominance, say
$x {\ \lower-1.2pt\vbox{\hbox{\rlap{$>$}\lower5pt\vbox{\hbox{$\sim$}}}}\ }0.1$.
 For  completely consistent analysis, the parameters $\alpha$ and $\beta$, and
 the distributions $\Delta u(x)$ and $\Delta d(x)$
should be scaled to the same value of $Q^2$. However,  
for the sake of the unsophisticated estimates we do not
go into such details.

The results of calculation of the nucleon  and deuteron 
 SFs, $h_1^N$ (solid lines) and $h_1^D$ (dashed lines),
are shown in   Fig.~5.
The group of curves 1 represents case (\ref{ab1}), which
is a possible lower limit   for   $h_1^{N,D}$
in accordance with
our estimates (\ref{ab3}). Curves 2 represent the case
  $\alpha = 1.5,\quad\beta = 0.5$, which is close
to the mid point results of the bag model and the QCD sum rules.
The upper limit corresponding to the estimates (\ref{ab3})
is presented by curves 3. For all cases
the deuteron
 SF is suppressed comparing to the nucleon one, mainly
because of the depolarization effect of the D-wave in the deuteron.
This is quite similar to the case of the SFs $g_1^N$
and $g_1^D$.  
Note that our  estimate of the nucleon SF $h_1^N$,
(\ref{ab3}),
gives systematically larger   function than naive suggestion (\ref{ab1}),
the curves 1 in Fig.~5
which essentially corresponds to the estimate $h_1^N \simeq (18/5) g_1^N$,
neglecting possible negative contribution of the s-quark sea~\cite{jaffer,rev2}. 
The large size of the effect suggests that it can be detected in 
future experiments
with the deuterons~\cite{exp}.
 
 \section{Brief conclusion}

We have presented the results of our study of the spin-dependent structure
functions of the
deuteron. In particular, the leading twist  $g_1^D$, $b_{1,2}^D$ and
$h_1^D$ are considered.
The issue of the extraction of the neutron structure functions 
from the deuteron data is addressed. The  role of relativistic effects
is studied and can be summarized as: (i) relativistic calculations
give a slightly larger magnitude of the binding effects, (ii) the 
relativistic Fermi motion results in ``harder'' SF
at high $x$, and (iii) covariant approach
is  internally consistent, while the nonrelativistic approach
is internally inconsistent and violates
important sum rules.

\section*{Acknowledgements}

We wish to thank every one who essentially contributed to studies
included in this presentation, L.P. Kaptari, C. Ciofi degli Atti, Han-xin He and
S. Scopetta.
This work is supported in part by   NSERC, Canada, and INFN, Italy. 


{\small
  \begin{thebibliography}{50} 

\bibitem{jaffer} R.L. Jaffe, e-print archives: hep-ph/9603422 and hep-ph/9602236. 

\bibitem{rev2} Hai-Yang Cheng, 
e-print archive: hep-ph/9607254.

\bibitem{amb} L.P. Kaptari and A.Yu. Umnikov, Phys. Lett. {\bf B259} (1991) 155.

\bibitem{unfo}   A.Yu. Umnikov, F.C. Khanna and L.P.Kaptari,  
Z. Phys. {\bf A348} (1994) 211.

\bibitem{mtamb} W. Melnitchouk and A.W. Thomas,
 Phys. Lett. {\bf B377} (1996) 11.

\bibitem{kubj} L.P. Kaptari and A.Yu. Umnikov, Phys. Lett. {\bf B240} (1990)   203.

\bibitem{fshe} L. Frankfurt, V. Guzey, M. Strikman, Phys.Lett.B381:379-384,1996. 


\bibitem{uk} A.Yu.  Umnikov 
 and F. Khanna, Phys. Rev. {\bf C49} (1994) 2311;\\
 A.Yu.  Umnikov, L.P.  Kaptari, K.Yu. Kazakov
 and F. Khanna, Phys. Lett. {\bf B334} (1994) 163.


\bibitem{ukk} 
A.Yu.  Umnikov, F. Khanna and L.P.  Kaptari, e-print archives:
hep-ph/9608549; {\em to be published}.


\bibitem{mst} W. Melnitchouk, A.W. Schreiber and A.W. Thomas, Phys. Rev.
{\bf D49} (1994) 1183.

\bibitem{kaemp}   L.P. Kaptari, K.Yu. Kazakov, A.Yu. Umnikov and 
B. K\" ampfer, Phys. Lett. {\bf B321} (1994)  271.

\bibitem{mg1}  W. Melnitchouk, G. Piller and A.W. Thomas,
 Phys. Let. {\bf B346}, 165 (1995).

\bibitem{expgdh} S.E. Kuhn et al., CEBAF Proposal No. 93-009;
 Z.E. Meziani et al., CEBAF Proposal No. 94-010.

\bibitem{res} C. Ciofi degli Atti, L.P. Kaptari, 
 S. Scopetta and  {A.Yu. Umnikov},
Phys. Lett. {\bf B376} (1996) 309.                  

\bibitem{ciofi} C. Ciofi degli Atti, S. Scopetta, e-print archive: nucl-th/9606034; 
{\em to be published}. 

\bibitem{ub} A. Yu. Umnikov, e-print archive: nucl-th/9605291.


\bibitem{et} A.V. Efremov and O.V. Teryaev, Sov. J. Nucl. Phys. {\bf 36}, 557 
(1982).


\bibitem{jj} R.L. Jaffe and X. Ji, Phys. Rev. Lett. {\bf 67} (1991) 552; 
            Nucl. Phys. {\bf B375} 527.

\bibitem{hj} H.X. He and X. Ji, Phys. Rev. D {\bf 52} (1995) 2960;\\
 H.X. He and X. Ji,  MIT-CTP-2551; e-print archive:  hep-ph/9607408.

\bibitem{exp} The RSC proposal to RHIC, 1993;
              The HERMES proposal to HERA, 1993.

\bibitem{uhk} 
A.Yu.  Umnikov, Han-xin He and F. Khanna, e-print archives:
hep-ph/9609353; {\em to be published}.



\bibitem{shaf} A. Sch\"afer, Phys. Lett. {\bf B208}, 175 (1988).






\end{thebibliography} 

}
\end{document}